\documentclass[prl,twocolumn,showpacs,preprintnumbers,amsmath,amssymb,superscriptaddress]{revtex4-1}

\usepackage{amsmath,amssymb,amsfonts} 	
\usepackage{graphicx}
\usepackage[latin1]{inputenc}
\usepackage{subfigure}

\renewcommand{\[}{\begin{equation}}
\renewcommand{\]}{\end{equation}}
\def\bea{\begin{eqnarray}}
\def\eea{\end{eqnarray}}
\def\nn{\nonumber\\}

\newcommand{\intr}{\int d{\bf r} \;}
\newcommand{\emi}[1]{{\rm e}^{-i #1}}

\newcommand{\p}{{\bf p}}

\newcommand{\kk}{\mbox{\boldmath$\kappa$}}
\renewcommand{\r}{{\bf r}}

\newcommand{\da}{\partial_{k_\alpha}}
\newcommand{\db}{\partial_{k_\beta}}

\newcommand{\equ}[1]{Eq.~(\ref{#1})}
\newcommand{\eqs}[2]{Eqs.~(\ref{#1}) and (\ref{#2})}

\def\bra#1{\langle#1\vert}
\def\ket#1{\vert#1\rangle}
\def\ev#1{\langle#1\rangle}

\def\runtime{(\the\time)\qquad\the\month/\the\day/\the\year}
\def\today
 {\count10=\year\advance\count10 by -2000 \number\day--\ifcase
  \month \or Jan\or Feb\or Mar\or Apr\or May\or Jun\or
             Jul\or Aug\or Sep\or Oct\or Nov\or Dec\fi--\number\count10}

\def\hour{\count10=\time\count11=\count10
\divide\count10 by 60 \count12=\count10
\multiply\count12 by 60 \advance\count11 by -\count12\count12=0
\number\count10 :\ifnum\count11 < 10 \number\count12\fi\number\count11}

\begin{document}

\title{Metal-insulator transition in disordered systems from the one-body density matrix}

\author{Thomas Olsen}
\email{tolsen@fysik.dtu.dk}
\affiliation{Centro de F{\'i}sica de Materiales, Universidad del Pa{\'i}s Vasco, 20018 San Sebasti{\'a}n, Spain}
\affiliation{Center for Atomic-Scale Materials Design, Department of Physics, Technical University of Denmark}

\author{Raffaele Resta}
\email{resta@democritos.it}
\affiliation{Dipartimento di Fisica, Universit{\`a} di Trieste, 34127 Trieste, Italy}
\affiliation{Donostia International Physics Center, 20018 San Sebasti{\'a}n, Spain}

\author{Ivo Souza}
\email{ivo_souza@ehu.es}
\affiliation{Centro de F{\'i}sica de Materiales, Universidad del Pa{\'i}s Vasco, 20018 San Sebasti{\'a}n, Spain}
\affiliation{Ikerbasque Foundation, 48013 Bilbao, Spain}

\begin{abstract}
The insulating state of matter can be probed by means of a ground state geometrical marker, which is closely related to the modern theory of polarization (based on a Berry phase). In the present work we show that this marker can be applied to determine the metal-insulator transition in disordered systems. In particular, for non-interacting systems the geometrical marker can be obtained from the configurational average of the norm-squared one-body density matrix, which can be calculated within open as well as periodic boundary conditions. This is in sharp contrast to a classification based on the static conductivity, which is only sensible within periodic boundary conditions. We exemplify the method by considering a simple lattice model, known to have a metal-insulator transition as a function of the disorder strength and demonstrate that the transition point can be obtained accurately from the one-body density matrix. The approach has a general {\it ab-initio} formulation and can be applied to realistic disordered materials by standard electronic structure methods.
\end{abstract}


\maketitle \bigskip\bigskip

The metal-insulator transition---either induced by electron-electron interaction (Mott transition) or by disorder in independent-electron  systems (Anderson transition)---has been studied by a variety of computational probes. In the Anderson case, the probes are invariably specific to model lattice Hamiltonians \cite{Kramer93}. Here we adopt a different and more general approach, stemming from the 1964 seminal paper by W. Kohn \cite{Kohn64,Kohn68}: 
according to Kohn the qualitative difference between insulators and conductors manifests itself in a different organization of the electrons in their many-body {\it ground state}. A series of more recent papers \cite{rap107,Souza00,rap132,rap_a31} has established Kohn's pioneering viewpoint on a sound formal and computational basis, rooted in geometrical concepts. These developments followed (and were inspired by) the modern theory of polarization, based on a Berry phase \cite{King93,rap_a28}. We will refer to these developments altogether as to the modern theory of the insulating state (MTIS); its basic ingredient is the quantum metric tensor \cite{Provost80}.

Over the years the MTIS has been adopted to address the Mott transition induced by correlation by adopting either lattice models \cite{rap107,Wilkens01,Tamura14,Varma15} or first-principle Hamiltonians \cite{Stella11,Elkhatib15}; to the best of our knowledge it has never been adopted to investigate the Anderson transition in three-dimensional (3D) disordered samples. In the latter case, the tools currently in use focus on properties either of the spectrum or of the individual Hamiltonian eigenstates \cite{Kramer93}. We stress that instead---in the independent-electron case---the only ingredient of MTIS is the ground-state density matrix. 

Here we address a paradigmatic model: a tight-binding Hamiltonian on a 3D simple cubic lattice, with random onsite matrix elements. The Anderson transition for this model has been addressed in the previous literature by means of various tools \cite{Kramer93,MacKinnon81,Hofstetter94,Slevin99,Rodriguez11}. Here we show that---according to MTIS basic tenet---the ground-state density matrix of finite samples within ``open'' boundary conditions (OBCs) carries the information needed to detect the metal-insulator transition.

For the sake of simplicity we address isotropic systems only, whose scalar longitudinal conductivity is \[ \sigma(\omega) = \sigma'(\omega) + i\sigma''(\omega) ; \] the real and imaginary parts $\sigma'$ and $\sigma''$ obey Kramers-Kronig relationships. In a conductor the low-$\omega$ real part of $\sigma$ takes the general form \cite{Allen06} \[ \sigma'(\omega) = D \, \delta(\omega) + \sigma_{\rm reg}'(\omega) , \label{def} \] where $D$ is the Drude weight, and the regular part $\sigma_{\rm reg}'(\omega)$ may be non-vanishing for $\omega \rightarrow 0$. The nomenclature owes to the classical Drude theory in the dissipationless limit, where $D= \pi e^2 (n/m)$; $n$ is the carrier density and $m$ the corresponding mass. Taking into account the Kramers-Kronig relationships and \equ{def}, we may also rewrite \[ \sigma(\omega) = D \, \left[ \delta(\omega) + \frac{i}{\pi \omega} \right] +\sigma_{\rm reg}(\omega) , \label{def2}\] whence the alternative definition \cite{Kohn64,note1} \[ D = \pi \lim_{\omega \rightarrow 0} \omega \sigma''(\omega) . \]
The insulating behavior of a material implies both $D=0$ and $\sigma'_{\rm reg}(\omega) \rightarrow 0$ for $\omega \rightarrow 0$ at zero temperature, while in conductors one has either $D \neq 0$ (in pristine crystalline metals) or $\sigma'_{\rm reg}(0) \neq 0$.

The Kubo formulae provides the quantum-mechanical expression for $\sigma'_{\rm reg}(\omega)$, while instead $D$ is a ground-state property. In the special case of a pristine crystal at the independent-particle level $D$ measures the current due to freely accelerating electrons at the Fermi surface, while $\sigma_{\rm reg}(\omega)$ is due to interband transitions. Both terms in \equ{def2}, however, have a more general meaning and are well defined even for an interacting many-body system \cite{Scalapino92}. In either case a non-vanishing static conductivity requires periodic boundary conditions (PBCs) and the vector-potential gauge for the electric field. Indeed there cannot be any steady-state current in a finite crystallite within OBCs. The Kubo formulae for the conductivity is the standard approach to discriminating between insulating and metallic phases. However, the MTIS implies that an alternative approach is possible as will be shown below. Notably, the difference between an insulator and a metal can be detected within either PBCs or OBCs. We will adopt the latter in the present investigation, stressing the fact the the metallic/insulating behavior is a \textit{ground state} property that can be adressed without reference to the static conductivity.

Consider $N$ interacting electrons in a box of volume $V$, with Hamiltonian (in atomic units) \[ \hat{H}(\kk) =  \frac{1}{2} \sum_{i=1}^N (\hat\p_i + \kk)^2 + \hat{U} , \label{hamiltonian} \] where $\hat{U}$ comprises one- and two-body interactions. At $\kk=0$ \equ{hamiltonian} is the standard many-body Hamiltonian of the system, while setting $\kk \neq 0$ amounts to a gauge transformation. Such a transformation within OBCs is trivial, and can be easily ``gauged away'': for instance, the ground-state energy is $\kk$-independent. Matters are instead nontrivial within PBCs, where the ground-state energy $E_0(\kk)$ is in general $\kk$-dependent. For the sake of clarity we remind that PBCs means that the wavefunction at any $\kk$ is periodical in the supercell of volume $V$ in each electronic coordinate (the coordinates are indeed angles). It has been shown by Kohn \cite{Kohn64,note1} that within PBCs the Drude weight is given (for isotropic systems) by \[ D = \left. \frac{\pi}{V} \frac{d^2 E_0(\kappa)}{d \kappa^2} \right|_{\kappa=0} . \]

If we define the projector \[ \hat{Q}(\kk)= \hat{1} - \ket{\Psi_0(\kk)} \bra{\Psi_0(\kk)} , \] the quantum metric tensor \cite{Provost80} is \[ G_{\alpha\beta}(\kk) = \frac{1}{N} \mbox{Re } \ev{\da \Psi_0(\kk) | \, \hat{Q}(\kk) \,| \db \Psi_0(\kk) } \label{metric}, \] where we have divided by $N$ in order to obtain an {\it intensive} quantity. This tensor has the dimensions of a squared length, and is a scalar in isotropic systems, where we define the MTIS localization length as \[ \lambda^2 = G_{\alpha\alpha}(0) , \label{lambda} \] in the thermodynamic limit. We note in passing that the imaginary part of $\langle\Psi_0(\kk) | \, \hat{Q}(\kk) \,| \db \Psi_0(\kk)\rangle$ is closely related to the Berry curvature of the system, thus emphasizing the geometric interpretation of the MTIS localization length. The MTIS basic tenet is that $\lambda$ is the main marker for the insulating state of matter: in fact $\lambda$ is finite in any insulator, while it diverges in any metal \cite{rap107,Souza00,rap132,rap_a31}. For the sake of clarity, we stress that the MTIS localization length $\lambda$ bears {\it no relationship} to the Anderson localization length \cite{Kramer93}: the former is a property of the many-body ground state, while the latter is a property of the one-body eigenstates in an independent-electron system. In the Supplementary material we demonstrate the relationship between the $\lambda$ and the regular part of the conductivity from which it follows that a finite static regular conductivity implies a diverging MTIS localization length.

The convergence/divergence of $\lambda$ has been often used to address the Mott transition in correlated systems \cite{rap107,Wilkens01,Stella11,Tamura14,Varma15,Elkhatib15}; the present Letter is about adopting the same viewpoint to address the Anderson transition in a 3D disordered system. The metal-insulator transition in presence of both disorder and electron-electron interaction has received much interest as well  \cite{Basko06}. Here we only quote two very recent simulations based on 1D model Hamiltonians within PBCs: Ref. \cite{Varma15} adopts MTIS, while Ref. \cite{Bera15} proposes a marker based on the one-body density matrix $\rho$. The two approaches are not equivalent, since in the correlated case $\lambda$ cannot be evaluated from a knowledge of $\rho$ only.

One of the virtues of the MTIS is that \eqs{metric}{lambda} can be equally well implemented within either PBCs or OBCs. In this work we adopt OBCs, where the metric assumes a very transparent meaning. If we define the many-body operator \[ \hat{\r} = \sum_{i=1}^N \hat\r_i , \] then the $\kk$-dependence of the ground eigenstate is very simple within OBCs: \[ \ket{\Psi_0(\kk)} = \emi{\kk \cdot \hat{\r}}\ket{\Psi_0(0)} = \emi{\kk \cdot \hat{\r}}\ket{\Psi_0}, \] with an obvious simplification of notations. From this we easily get \bea  \left. \da  \ket{\Psi_0(\kk)} \; \right|_{\kk = 0} = -i \hat{r}_\alpha \ket{\Psi_0} \\ G_{\alpha\beta}(0) =\frac{1}{N} \mbox{Re }( \, \ev{\Psi_0 | \hat{r}_\alpha \hat{r}_\beta | \Psi_0 } \nn - \ev{\Psi_0 | \hat{r}_\alpha | \Psi_0} \ev{ \Psi_0 |\hat{r}_\beta | \Psi_0 } \, ) , \label{open} \eea i.e. the metric tensor is the second cumulant moment of the electron distribution in the many-electron system. From \equ{open} it is clear that within OBCs the MTIS localization length is a function of the {\it two-body} density matrix \cite{rap132}. In the case of noninteracting particles \equ{open} can be expressed in terms of the one-body density matrix as \[ G_{\alpha\beta}(0) = \frac{1}{2N} \intr \!\!\! \int d \r' (\r -\r')_\alpha (\r -\r')_\beta |\rho(\r,\r')|^2.  \label{cumul} \] Here we have adopted a ``spinless electron'' formulation, which we will use throughout the present work. The scaling behavior of $|\rho(\r,\r')|$ for $|\r - \r'| \rightarrow \infty$ determines whether the integral in \equ{cumul} converges or diverges in the large-system limit. The crystalline case is well known \cite{Ismail99}: $|\rho(\r,\r')|$ decays exponentially in insulators and algebraic in metals, resulting in convergence in the former case, and typically divergence in the latter.

In a disordered system $|\rho|^2$ in \equ{cumul} has to be replaced with its configurational average $\ev{|\rho^2|}_{\rm c}$. A very crucial point is that $\ev{|\rho^2|}_{\rm c}$ is in general {\it different} from the squared modulus of the configurational average of $|\rho|$. Thus, knowing the decay of $|\rho|$ is in general not sufficient to determine whether a disordered system is insulating or metallic. This is closely related to the so-called vertex corrections in the well established transport theories based on Green's functions \cite{Ziman,Allen06}. We discuss this point in detail in the Supplementary Material \cite{supplementary}.

Our case study is a paradigmatic system displaying the metal-insulator transition. We consider the half-filled  3D tight-binding model
\begin{align}
 H=t\sum_{<ij>}c^\dag_ic_j+\mbox{H.c.}+W\sum_i\varepsilon_ic^\dag_ic_i,
\end{align}
where $i,j$ denote sites on a simple cubic lattice, $<ij>$ are pairs of nearest neighbor sites and the onsite energies $\varepsilon_i$ are randomly picked from the interval $[-1,1]$. $W$ is the disorder strength and the model has previously been shown to exhibit an Anderson transition at $W_{\rm c}/t=8.25$ \cite{MacKinnon81,Hofstetter94,Slevin99,Rodriguez11}. We set $t=1$ in the following. 

\begin{figure}[b]
    \includegraphics[width=4.0 cm]{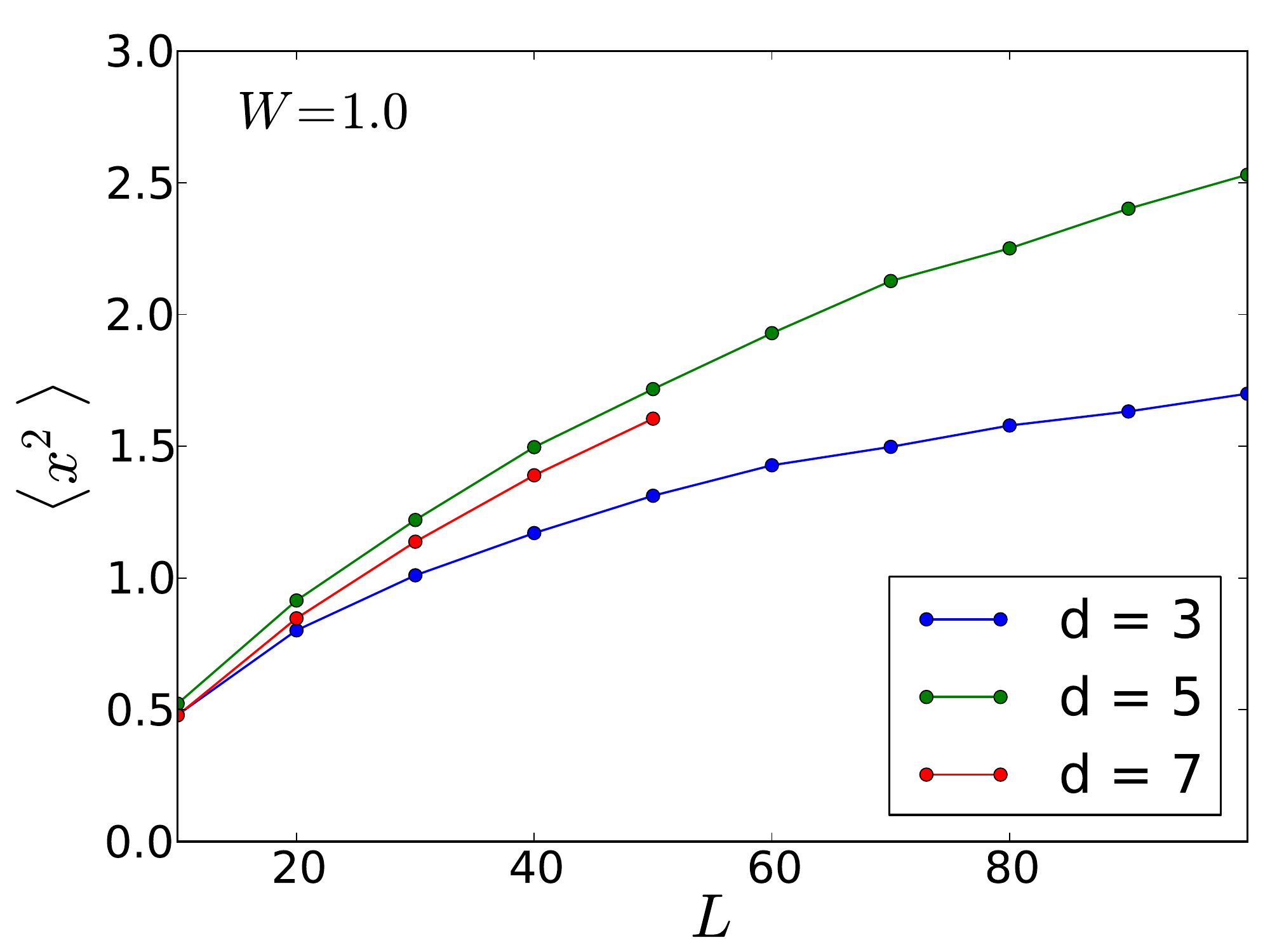}
    \includegraphics[width=4.0 cm]{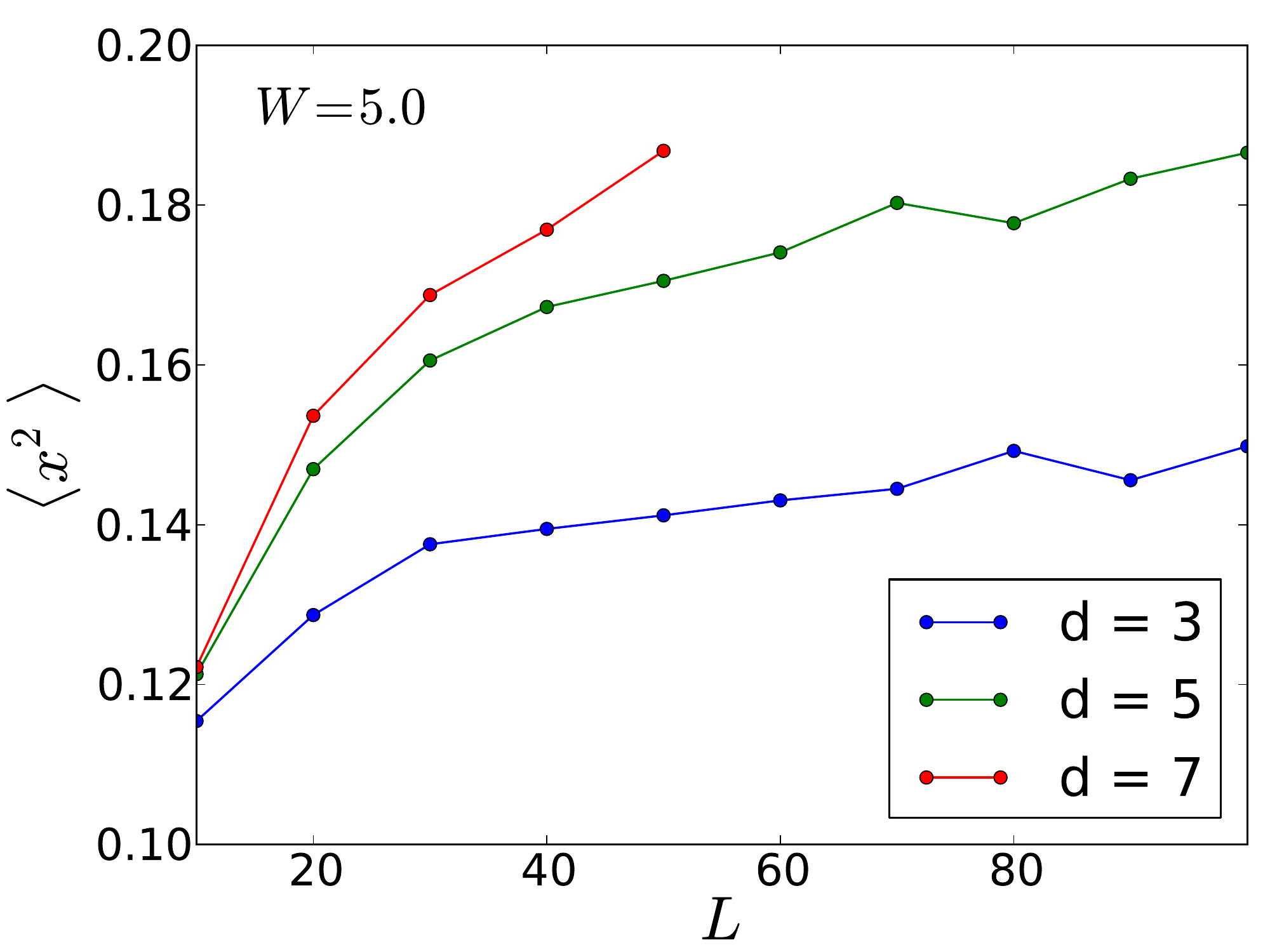}
    \includegraphics[width=4.0 cm]{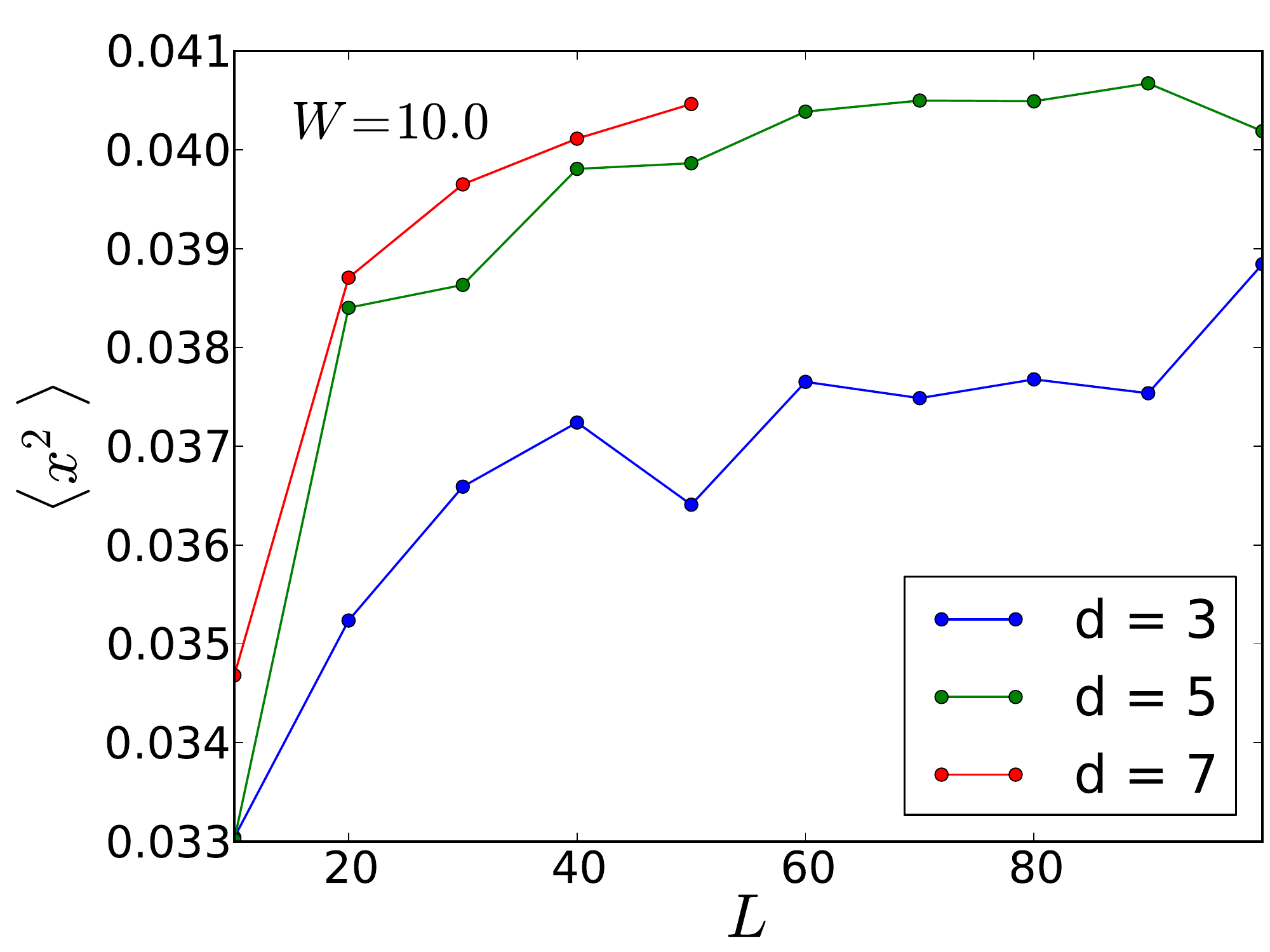}
    \includegraphics[width=4.0 cm]{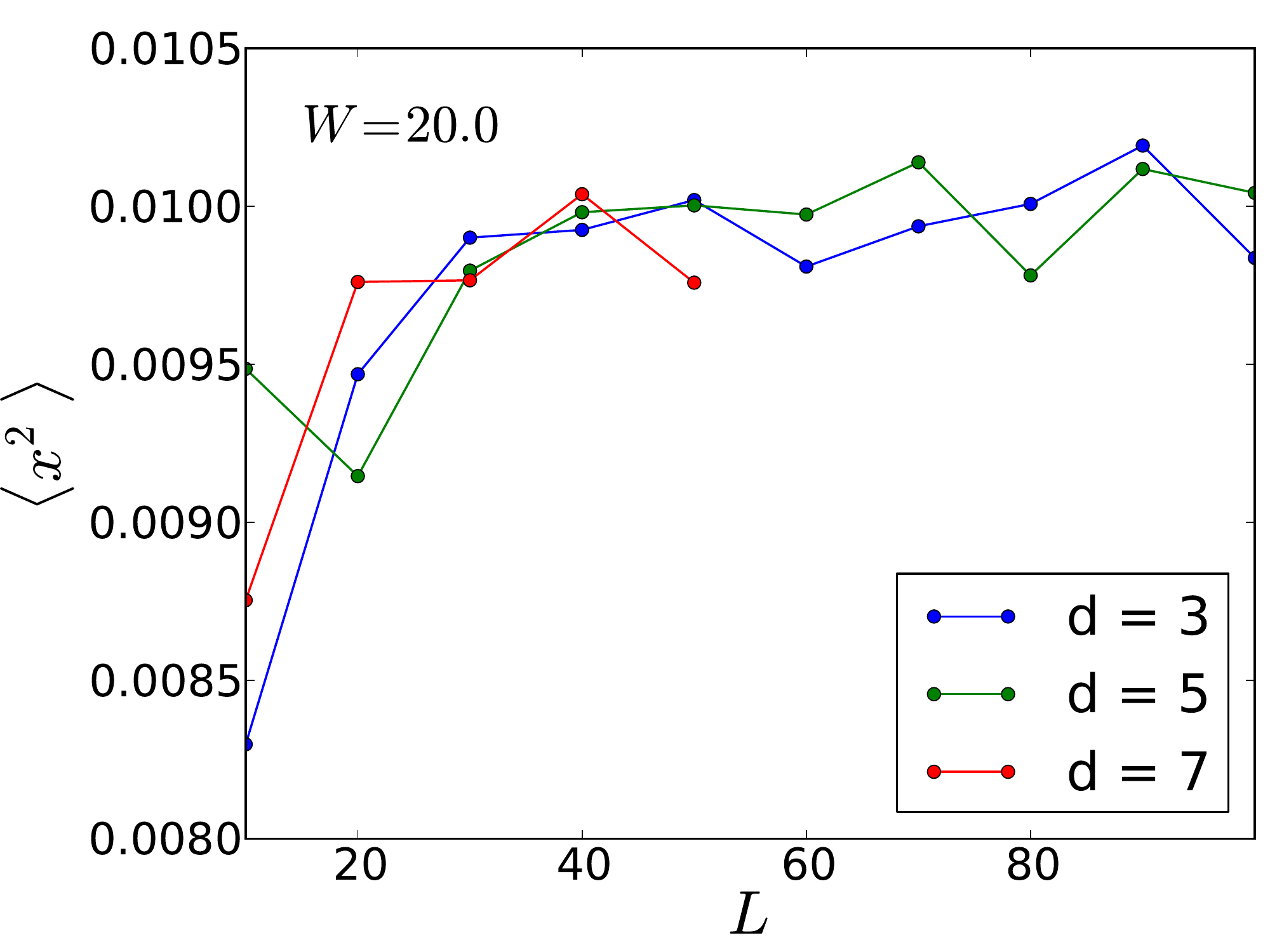}
\caption{(Color online). Localization length $\lambda$ as a function of rod length $L$. $\lambda$ diverges for small values of $W$ and saturates to a finite value for large values of $W$.}
\label{fig:x2}
\end{figure}

We have calculated the localization length $\lambda$, \equ{lambda}, within OBCs for various values of $W$ using rods of size $L\times d\times d$ where $L=100$ and $d=3,5,7$. To obtain the configurational average we used 100 configurations and for each configuration the component of the localization tensor, \equ{cumul}, along the rod was obtained by averaging over the two short dimensions. The results for various values of $W$ are shown in Fig. \ref{fig:x2} for different rod widths $d$. We clearly observe a tendency for  $\lambda$ to saturate when $W$ becomes large. For small $W$, instead, $\lambda$  appears to be increasing monotonically with the rod length $L$. Within MTIS the Anderson transition would emerge as a transition from a divergent to a finite $\lambda$ in the limit of large $L$. While it seems plausible that this may happen around $W_{\rm c}=8.25$, it is very difficult to extract a quantitative estimate of $W_{\rm c}$ from  $\lambda$ alone. For example, for $W=10$, the localization length appears to be saturated at a finite value for $L\sim100$, but it is hard to verify if this is really the case or if $\lambda$ is merely increasing too slowly to be observable at the size of our simulations.

\begin{figure}[t]
    \includegraphics[width=8.0 cm]{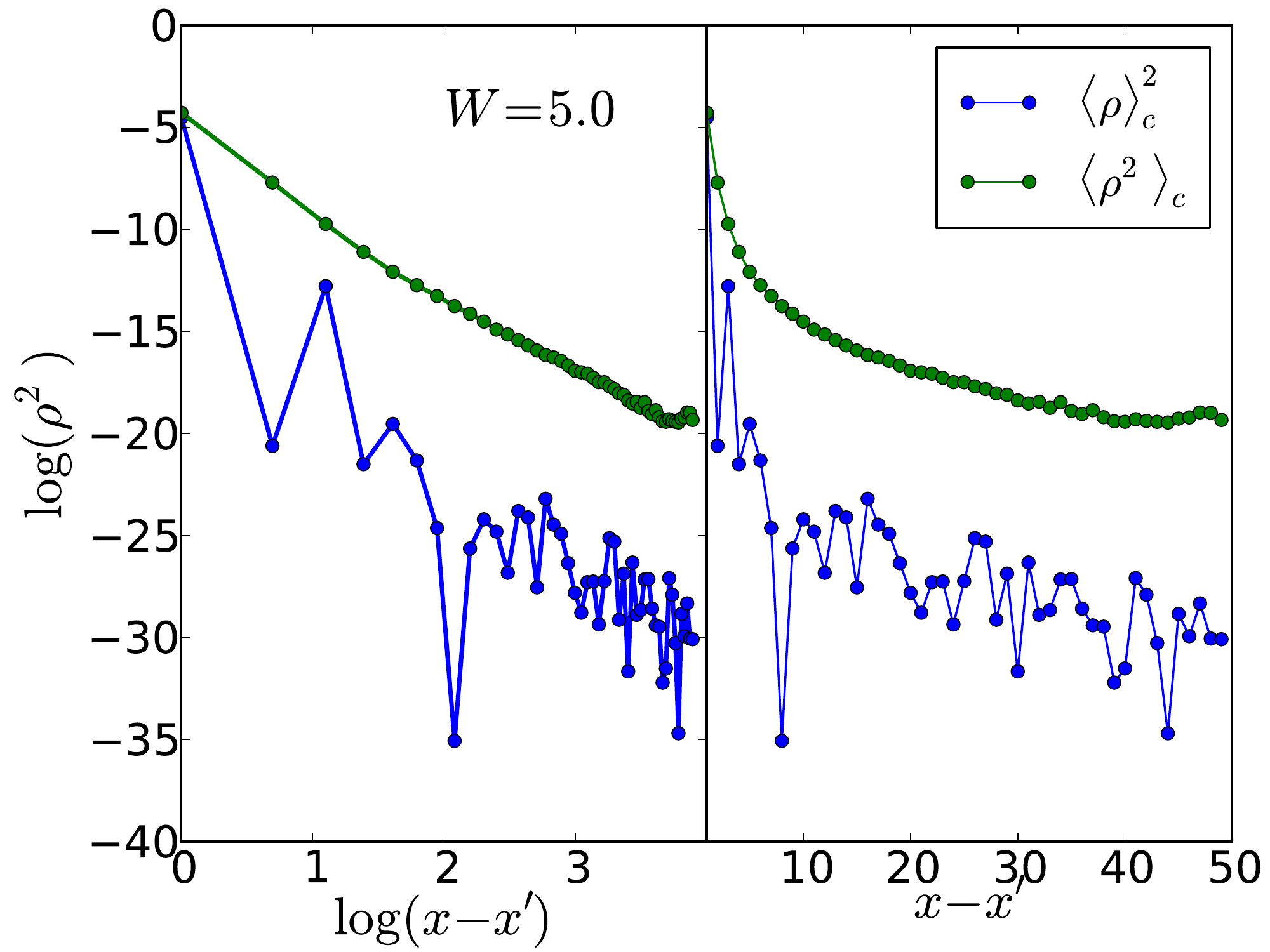}
    \includegraphics[width=8.0 cm]{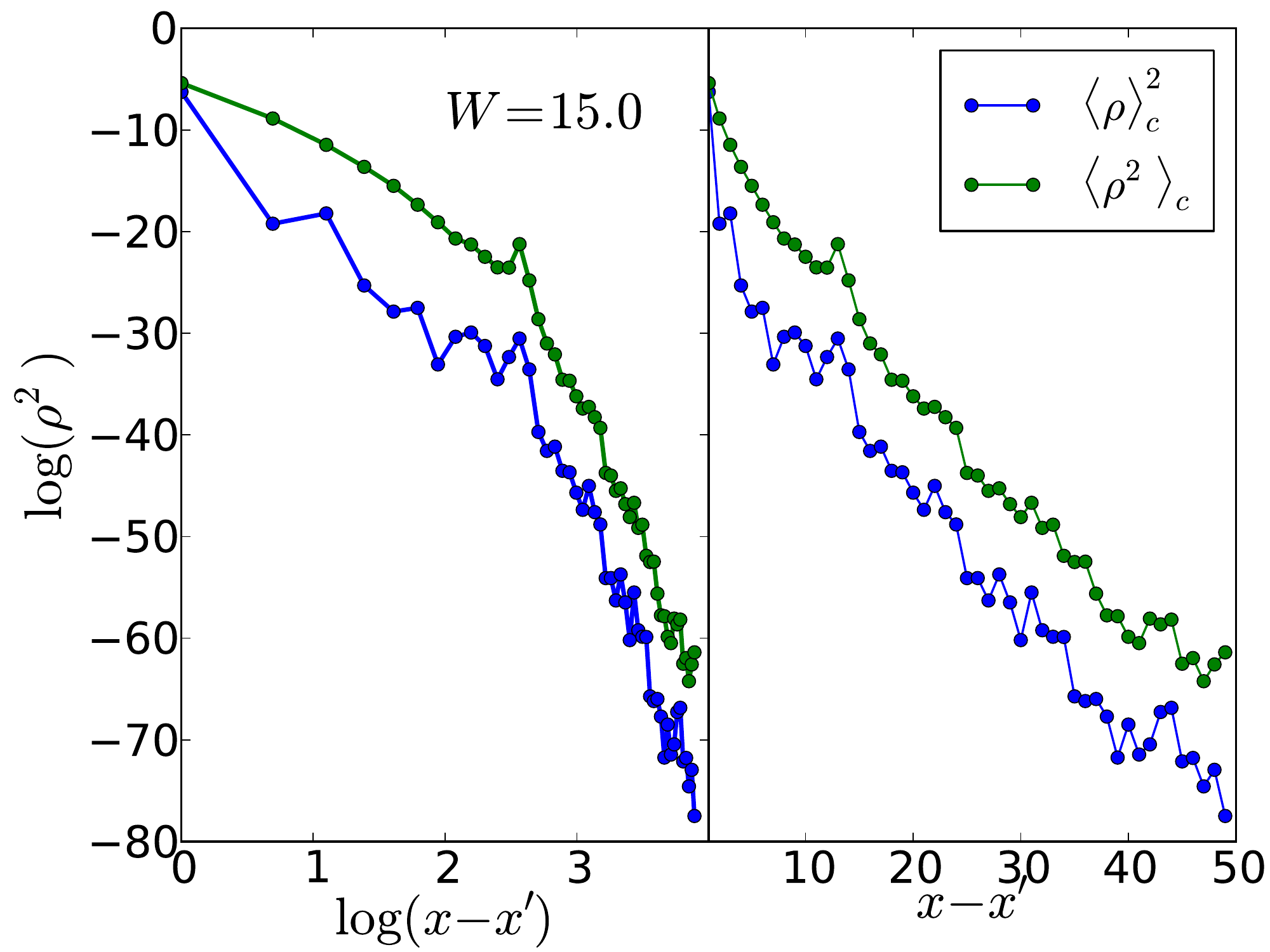}
\caption{(Color online). Configurational averaged density matrix. Top: density matrix with $W=5.0$ in log-log scale to the left and in semi-log scale to the right. Bottom: same as top, but with $W=15.0$. The norm-squared density matrix is seen to be well approximated by power-law decay for $W=5.0$ and exponential decay for $W=15.0$.}
\label{fig:rho}
\end{figure}
In the following we will analyze the density matrix directly, showing that the Anderson transition can be indeed detected from the long range behavior of $\ev{|\rho^2|}_{\rm c}$. As discussed above (and in the Supplementary Material) it is essential to take the square of the density matrix {\it before} the configurational average, and not the reverse. In Fig. \ref{fig:rho} we show the result of our computer experiments, performed for $W=5$ (in the conducting regime) and $W=15$ (in the Anderson-insulating regime), after averaging over 300 random configurations; both options---$\ev{|\rho^2|}_{\rm c}$ and $\ev{|\rho|_{\rm c}}^2$---are shown, and both are plotted in semi-logarithmic and double logarithmic scales. The panels in Fig. \ref{fig:rho} show first of all that $\ev{|\rho^2|}_{\rm c}$ is a much smoother quantity: this property will allow us (see below) to  locate the critical disorder strenght $W_{\rm c}$. The top left panel in Fig. \ref{fig:rho} clearly indicates a power-law behavior at $W=5$, while the bottom right panel indicates an exponential behavior at $W=15$: this is indeed qualitatively consistent with Fig. \ref{fig:x2}, and also with analytical results in the literature \cite{Aizenman98}. 
It should be noted, however, that exponential decay is a sufficient, but not a necessary condition for the finiteness of $\lambda$. For example, in a homogeneous system it can be seen from \equ{cumul} that $\lambda$ stays finite if $\ev{|\rho^2|}_{\rm c} \sim |\r - \r'|^{-\beta}$ and $\beta>5$.

\begin{figure}[tb]
    \includegraphics[width=8.0 cm]{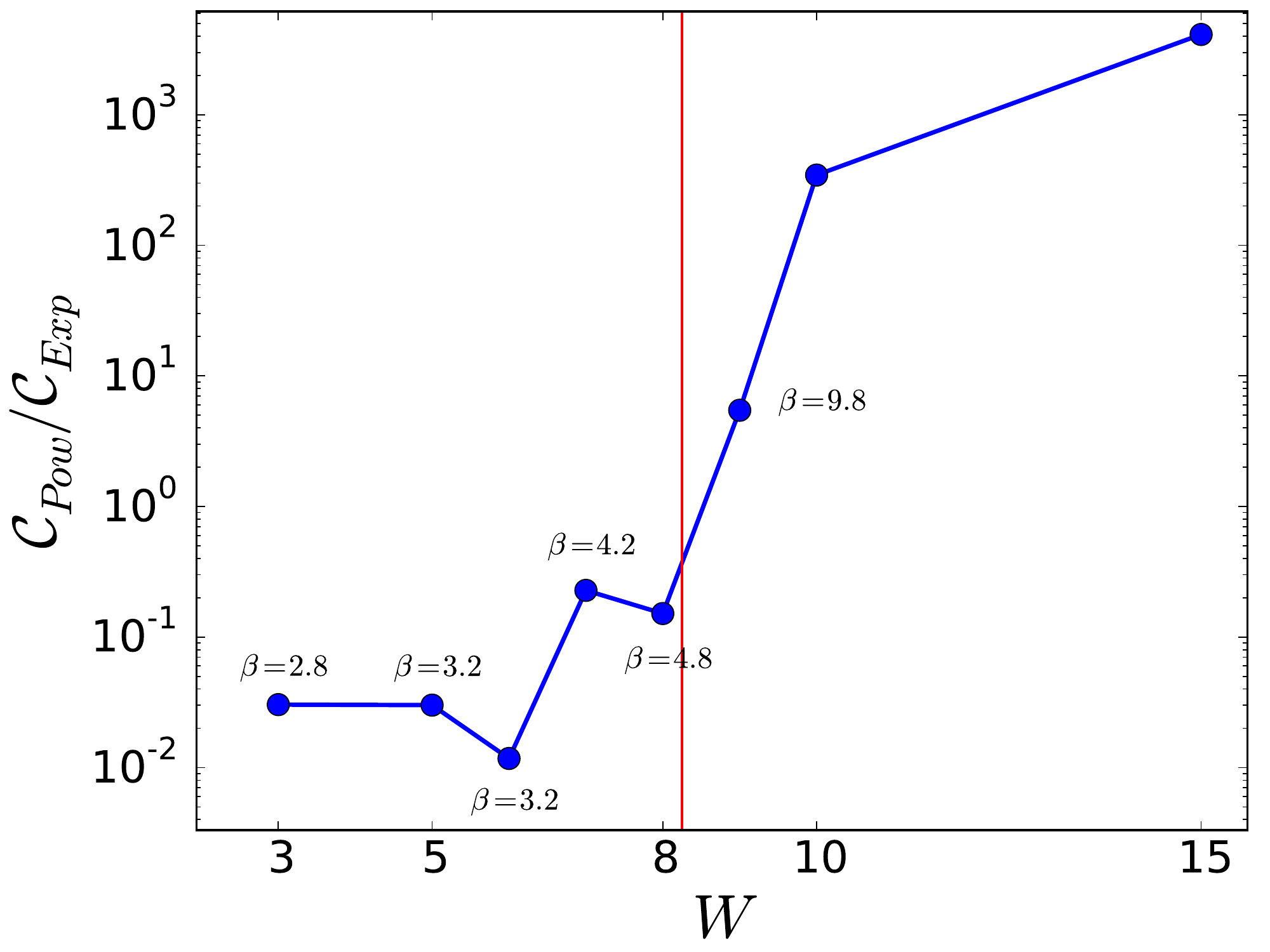}
\caption{(Color online). Ratio of the two cost functions, \equ{cost}, from a least-square fit using both power-law and exponential formulae. The displayed values of $\beta$ are the fitted exponents. The vertical red line is at the value $W_{\rm c}=8.25$, taken from the literature \cite{MacKinnon81,Hofstetter94,Slevin99,Rodriguez11}. Our best estimate of the metal insulator transition from the present method is where $\mathcal{C}_{\rm pow}/\mathcal{C}_{\rm exp}$ becomes unity. This happens at $W\approx8.5$.}
\label{fig:prob}
\end{figure}
In order to get a quantitative estimate for the Anderson transition, we consider two alternative formulae for representing the scaling of $y(x)=\langle|\rho(x)|^2\rangle_c$, where we set $x =  |\r - \r'|$. The two formulae have either power-law or exponential decay:
\begin{align}
 \tilde y_{\rm pow}(x)=ae^{-bx},\\
 \tilde y_{\rm exp}(x)=\alpha x^{-\beta}.
\end{align}
We indicate with $\tilde y_{\rm X}$ any of the two. Then, assuming constant Gaussian noise, the probability of obtaining the data displayed in Fig. \ref{fig:rho} using each of the two formulae is 
\begin{align}
P_{\rm X}\sim e^{-\mathcal{C}_{\rm X}} ,
\end{align}
where the ``cost'' function is
\begin{align}
\mathcal{C}_{X}=\sum_i\frac{(\tilde y_{\rm X}(x_i)-y_i)^2}{2\sigma^2}. \label{cost}
\end{align}
Here the index $i$ labels lattice sites along $L$ and  $y_i$ are configuration-averaged values of $\langle|\rho(x_i)|^2\rangle_{\rm c}$. 

We can then obtain the parameters in the two formulae by a least-square fit and compute the resulting cost function for either formula. In Fig. \ref{fig:prob} we show the cost-function ratio, as obtained from a fit to the two formulae: we observe a very steep increase (two orders of magnitude) between $W=8$ and $W=9$. The transition is therefore very sharp using our indicator, which switches from nearly vanishing to one in a narrow $W$ interval. The present approach yields a critical disorder parameter $W_{\rm c}\approx8.5$.
It should also be noted that the fitted exponents in the $W$ region where power-law decay is most likely satisfy $\beta<5$, i.e. all yield a divergent $\lambda$.

In conclusion we have proved that the modern theory of the insulating state, adopted so far in the previous literature for band insulators and Mott insulators, successfully applies even to a paradigmatic Anderson insulator. The standard computational methods to address the Anderson transition are often peculiar to lattice models (recursive methods and the like), while the MTIS approach adopted here is quite general and would apply to \textit{ab initio} studies as well. Furthermore, the general expression \equ{open} is valid for many-body systems and thus provides a general framework to include interactions in the study of the Anderson transition.

T.O. acknowledges support from the Danish Council for Independent Research, Sapere Aude Program; R.R.
acknowledges support from the ONR (USA) Grant No. N00014-12-1-1041; I.S. acknowledges support from Ministerio de Econom\`ia y Competitividad (Spain) Grant No. MAT2012-33720, and from the European Commission Grant No. CIG-303602.


\begin{thebibliography}{10}
\bibitem{note1}
{ We adopt the same normalization and signs as in Ref. \cite{Allen06}; this is
  different from Ref. \cite{Kohn64}}.
\end{thebibliography}
\end{document}